\documentstyle[preprint,revtex]{aps}

\begin{document}
\draft

\begin{title}
Fluctuations and pattern selection near an Eckhaus instability
\end{title}

\author{E. Hern\'andez-Garc\'\i a$^{1}$, Jorge Vi\~nals$^{2}$,
Ra\'ul Toral$^{1,3}$, and M. San Miguel$^{1}$}

\begin{instit}
$^{1}$Departament de F\'\i sica, Universitat de les Illes Balears \\
E-07071 Palma de Mallorca, Spain
\end{instit}

\begin{instit}
$^{2}$Supercomputer Computations Research Institute, B-186\\
Florida State University, Tallahassee, Florida 32306-4052
\end{instit}

\begin{instit}
$^{3}$Institut d'Estudis Avan\c cats de les Illes Balears, \\
Consejo Superior de Investigaciones Cient\'\i ficas and \\
Universitat de les Illes Balears, E-07071 Palma de Mallorca, Spain
\end{instit}
\begin{abstract}
We study the effect of fluctuations in the vicinity of
an Eckhaus instability. The classical stability limit, which is defined in
the absence of fluctuations, is smeared out into a region in which
fluctuations and
nonlinearities dominate the decay of unstable states.
The width of this region is shown to grow as $ D^{1/2} $,
where $D$ is the intensity of the fluctuations. We find an effective stability
boundary that depends on $D$.
A numerical solution of the stochastic Swift-Hohenberg equation in one
dimension
is used to test this prediction and to
study pattern selection when the initial unstable state lies within the
fluctuation dominated region. The asymptotically selected state differs from
the
predictions of previous analyses. Finally, the nonlinear
relaxation for $D > 0$ is shown to exhibit a scaling form.
\end{abstract}

\narrowtext

Recent experimental studies of Rayleigh-B\'enard convection
in ${\rm CO_2}$ gas \cite{re:bo91}, convection in binary mixtures
\cite{binary},
electro-hydrodynamic convection in nematic liquid crystals \cite{re:re91},
and Taylor-Couette flow \cite{fluctTC}, have detected random
fluctuations of thermal origin that are strong enough to be analyzed
quantitatively. Whereas the effect of thermal fluctuations on thermodynamic
equilibrium
is now well understood in general terms, much less is known about their effect
in the vicinity of nonequilibrium instabilities. Pioneering work in this
direction
was done by Graham \cite{re:gr74}, and Swift and Hohenberg \cite{re:sw77}.
It was believed, however, that since
characteristic thermal energies are several orders of magnitude smaller than
the characteristic driving or dissipative energies involved in many
nonequilibrium instabilities,
the effect of fluctuations would be far too small to be observable.
The results of these recent experiments have renewed interest in the subject
since they have opened the possibility of finding nontrivial phenomena due to
fluctuations close to nonequilibrium instability points.

We study in this paper the effect of fluctuations on the
Eckhaus instability \cite{re:ec65}. The Eckhaus instability is a longitudinal
instability often exhibited in
systems that display patterns which are spatially periodic.
It has been exhaustively studied in many
systems, including, for example, the stability of a set of parallel convective
rolls in Rayleigh-B\'enard convection \cite{re:ma90}, the stability of a
periodic array of cells during directional solidification
\cite{re:br90}, and instabilities of stationary standing
Faraday waves \cite{re:do89}. We also note particularly
detailed experimental studies of the Eckhaus instability in
electro-hydrodynamic convection in nematic liquid crystals by Lowe and
Gollub \cite{re:lo85}, and Rasenat {\it et al.} \cite{rasenat}.
Three main issues are addressed here: the
fluctuation-induced smearing of the classical Eckhaus boundary, the
nonlinear relaxation
of the unstable solution, and the asymptotic periodicity of the new
stable solution following the instability (this is often referred to as
pattern selection).

In the absence of fluctuations, the Eckhaus boundary of potential or gradient
systems is located at the line where both the first and second
functional derivatives of the appropriate Lyapunov functional vanish; hence it
separates regions of metastability and instability. This is the analog of a
spinodal line in a first order phase transition. Our study can be
motivated by
analogy with this latter case: if fluctuations are included in the description
of a first order phase transition, the spinodal line ceases to exist
\cite{gsms}. Instead, states that are not stable are classified as metastable
or unstable with reference to a particular temporal
scale of evolution (observation time).
States that do not decay within such a scale are said to be
metastable, and unstable otherwise.
The often narrow region that separates both
types of behavior is shifted with respect to the spinodal line, and defines
an effective stability boundary known as
the \lq\lq cloud point". We found here related phenomena, namely
the smearing of the Eckhaus
boundary for finite amplitudes of the fluctuations, and the emergence of a
different time scale for the decay of the unstable state. This scale
depends on the intensity
of the fluctuations and identifies the observable stability
boundary. Our analysis of the fluctuation-dominated regime is similar in
spirit to Binder's determination of the
range of validity of a linearized description of phase
separation \cite{re:bi84}. He derived the
analog of a Ginzburg criterion to describe the effect of fluctuations on
the long wavelength instability known as spinodal decomposition.
Finally, we also discuss pattern selection when the initial
state is within the transition region.

Although we expect our analysis to be of wider applicability, we restrict
ourselves here to the one dimensional stochastic Swift-Hohenberg
(sSH) equation \cite{re:sw77},
\begin{equation}
\label{eq:sh}
\frac{\partial \psi(x,t)}{\partial t } = \left[ \epsilon - \left( 1 +
\frac{\partial^{2}}{\partial x^{2}} \right)^{2} \right] \psi(x,t)-
\psi(x,t)^{3}
+ \sqrt{D}\xi(x,t)
\end{equation}
which is known to be a model of Rayleigh-Be\'nard convection near onset. The
periodicity of the stationary solutions of Eq.(\ref{eq:sh})
is commensurate with the convective rolls, $\epsilon$
is the reduced Rayleigh number and $\xi(x,t)$ is a Gaussian random process
satisfying,
\begin{equation}
\label{eq:noise}
\left< \xi(x,t) \right> = 0, ~~~ \left< \xi(x,t) \xi(x',t') \right> = 2
\delta (x - x') \delta (t - t').
\end{equation}
The value of $D$ is proportional to the intensity of the thermal fluctuations
in the fluid. For $D=0$ and $\epsilon > 0$, Eq. (\ref{eq:sh}) admits periodic
stationary solutions $\psi_{q} (x)$ of wavenumber $q$ with
$q_{-L} < q < q_{L}$, and
$q_{\pm L} = \sqrt{1 \pm \sqrt{\epsilon}}$ \cite{re:maxx}. However,
only those solutions
with $q_{-E} < q < q_{E}$ ($q_{-L}
< q_{-E} <q_{E}< q_L$) are linearly stable.
Periodic solutions with $q$ outside this range exhibit a
modulational instability known as the Eckhaus instability.

The decay of the unstable solution is triggered by fluctuations.
In order to study such decay, we linearize
Eq. (\ref{eq:sh}) around a stationary periodic solution $\psi_{q_i}(x)$,
with $q_{E} < q_{i} < q_{L}$.
Linearization leads to an eigenvalue problem for a
spatially periodic operator. According to Floquet's
theorem, the normalized eigenfunctions $f_{q_i,k}(x)$ are of the form
\cite{bloch},
\begin{equation}
\label{bloch}
f_{q_i,k}(x) = \sum_{n=-\infty}^\infty \omega^{n}_{q_{_i},k}
e^{i(nq_i + k)x} \ ,
\end{equation}
each associated with an eigenvalue $\lambda(q_i,k)$, which determines
the growth rate of that particular eigenfunction. Because the SH
equation is real, these eigenvalues
satisfy the symmetry relation $\lambda(q_i,-k)=\lambda(q_i,k)$. Also,
the invariance of the SH equation under spatial translations implies
$\lambda(q_i,k=0)=0$. Let $\pm k_m$ be the values of $k$ that maximize
$\lambda(q_i,k)$ and let $f_m(x)$ be the real (and normalized)
combination of their associated eigenfunctions. Then, it
is possible to describe the early stages of the decay of the unstable state
by a linear stochastic equation governing the amplitude $u(t)$ of $f_{m}(x)$,
in the eigenfunction expansion of $\psi(x,t)-\psi_{q_i}(x)$:
\begin{equation}
\label{linear}
\dot u(t) = \lambda_m u(t) + \sqrt{D}\eta(t) \ .
\end{equation}
$\eta(t)$ is a white Gaussian random process which
results from the projection of $\xi(x,t)$ onto $f_m(x)$:
$\eta(t) = \int dx ~ f_m(x) \xi (x,t)$, the average of which is zero.
When the eigenfunctions in Eq. (\ref{bloch}) are normalized, the
variance of $\eta$ turns out to be independent of the initial
periodicity $q_i$.
Equation (\ref{linear}) is valid until a time $t_p$ at which the
amplitude $u$ becomes large enough such that nonlinearities
become important.
The solution of Eq. (\ref{linear}) with initial condition $u(t=0)=0$ is
$u(t) = h(t) e^{\lambda_m t}$, with
\begin{equation}
\label{ht}
h(t) = \sqrt{D}\int_0^t ds\ e^{-\lambda_m s} \eta(s) \ .
\end{equation}
The amplitude $h(t)$ is a Gaussian random process, averaging to zero, and
with a
time-dependent variance. From its definition it is easy to see that it becomes
stationary after a time $t_{m} \sim \lambda_m^{-1}$. After this time,
one can replace $h(t)$ by its long time limit,
$h(t=\infty)$, which is a Gaussian variable of standard deviation
$\sqrt{\left< h^2 \right>} = \sqrt{ {D / \lambda_m}}$.
Within this approximation, the calculation of the average time $t_p$ at
which $u(t)$ crosses a given reference value $u_T$ is standard
and gives \cite{mfpt},
\begin{equation}
\label{passage}
t_p = {1 \over \lambda_m} \ln \left( u_T \sqrt{\lambda_m \over D} \right)\ .
\end{equation}
Two characteristic time scales emerge from this analysis. The time $t_m$
signals the end of the fluctuation dominated regime and the beginning of a
linear deterministic regime. By choosing $u_T$ equal to a given (small)
fraction of
the final saturation value of $u$, $t_p$ can be interpreted as the time at
which nonlinear terms begin to be important. The approximation of replacing
$h(t)$ by its
asymptotic value is obviously true only if $t_m \ll t_p$, which implies
$u_T^2 \lambda_m \gg D$. When
\begin{equation}
\label{quasiginzburg}
\lambda_m \approx {D \over u_T^2}
\end{equation}
there is no clear
separation of time scales, and there will not be a distinct
linear deterministic
regime. Nonlinear effects then become important even in the early fluctuation
dominated regime.

The arguments give above are of a general nature. Consider now an initially
periodic state of wavenumber $q_{i} \agt q_E$.
The linear growth rate of the most
unstable eigenfunction vanishes near the Eckhaus boundary as
\begin{equation}
\label{wcritical}
\lambda_m \sim (q_i - q_E)^2.
\end{equation}
This can be seen by noting that the symmetry properties of
$\lambda(q_i,k)$ stated above imply
that, for small $k$, $\lambda(q_i,k) \approx a k^2 + b k^4 + ...$,
where $a$ and $b$ are functions of $q_i$ and $\epsilon$. $b$ must be positive
to avoid instabilities with arbitrarily large wavenumber, and $a$
changes sign at the Eckhaus boundary $q_i=q_E$. Then, to first order,
$a \sim (q_i - q_E)$. The maximum
$\lambda_m(q_i)$ of $\lambda(q_i,k)$ is at $k_m=\pm \sqrt{a/2b}$,
which leads to Eq. (\ref{wcritical}).
In addition, it can be seen that the
most important wavenumber in the Fourier expansion of the
eigenfunction $f_m(x)$ is $q_m\equiv
q_i-k_m$. From the expression for $k_m$ we get
\begin{equation}
\label{qM}
q_i - q_m  \sim (q_i - q_E)^{1 \over 2}.
\end{equation}

We now define a transition region of initial wavenumbers $q_{i}$ around
$q_{E}$, within which the extent of the linear deterministic regime vanishes.
{}From Eqs. (\ref{quasiginzburg}) and (\ref{wcritical}) we find that this
transition region is determined by
\begin{equation}
\label{ginzburg}
0 \le q_i - q_E  \le \frac{C}{u_T} \sqrt{D},
\end{equation}
where $C$ is a constant of order one.
Note that $u_{t}$ is finite near the Eckhaus
boundary, since it has to be taken as a finite fraction of the final saturation
amplitude of the fastest growing mode, and this quantity has no singularity
across the Eckhaus boundary.

These
results are restricted to an initially unstable state; hence they
only apply to $q_{i} > q_{E}$.
Initial states with $q_{i} < q_{E}$ which are
linearly stable according to a deterministic calculation, can also decay
because
of fluctuations, extending the transition region to $q_{i} < q_{E}$.
The smearing of the Eckhaus boundary and the
concomitant reduction of the region of linearly stable states can be thought of
as an effective shift of the Eckhaus boundary to a new value $\tilde{q}_{E} <
q_{E}$. It should be stressed, however, that this new effective boundary is
only defined with reference to a given observation time for the decay of
the metastable state.

We next address the implications of our analysis on the issue of pattern
selection. Pattern selection refers to the determination of a global
wavenumber $q_{f}$ of a configuration obtained after a long time, as a
function of the initial wavenumber $q_{i}$. The wavenumber $q_{f}$ is defined
here as $q_{f} = \pi n_{f}$, where $n_{f}$ is the number of nodes of the
configuration per unit length \cite{re:fo1}.
If $q_{i}$ lies outside the transition region, the initial stages of
the decay of the unstable state are well described by linear theory,
and the fastest growing
mode in the linear regime, $q_{m}$, will mostly determine $q_{f}$.
In this case, the evolution does not lead in observable time scales to a
periodic configuration of wavenumber $q_{min}$ that minimizes the Lyapunov
functional associated with Eq. (\ref{eq:sh}). In fact,
earlier numerical simulations of the sSH equation showed that
$q_{f} \approx q_{m}$ when $q_{i}$ is not too close to the Eckhaus boundary
\cite{estella}.
These results were consistent with previous simulations of the amplitude
equation also in the absence of noise \cite{kramer2}.
On the other hand, if $q_i$ lies inside the transition
region, nonlinearities and fluctuations are likely to alter these
conclusions.

To provide test our predictions we carried out a
computer simulation study of the one-dimensional
sSH equation. Complete details of the algorithm used can be found
elsewhere \cite{ours}. We chose $\epsilon = 0.05625$ and
discretized Eq. (\ref{eq:sh}) on an
evenly spaced grid with $N=8192$ nodes and $\Delta x = 2 \pi / 32$.
The equation was integrated forward in time with $\Delta t =
1 \times 10^{-4}$, and the
results averaged over a number of independent runs, typically 10-20. With this
choice of $\Delta x$, $\Delta q= 2\pi/ N
\Delta x = 0.004$.

Figure 1 presents the average value of $q_{f}$ as a
function of $q_{i}$ for
$D/ \Delta x = 0.05, 0.1$ and $0.2$.  The dashed line is
the value of $q_m$ found by numerically calculating the Floquet spectrum
on the same grid used to solve Eq. (\ref{eq:sh}). The analysis also
yields $q_E \approx 1.201$.
Simulation points for $q_i < q_E$ are also shown in
Fig. ~\ref{master}: although in the stable range according to the deterministic
calculation, they are seen to decay. Furthermore,
for $q_{i}$ far enough from the Eckhaus boundary the value of $q_{f}$
obtained from the simulation
approaches $q_m$. However as $q_i - q_E$ is decreased, the data deviate
from $q_m$. The size of the region where $q_f$ deviates from $q_m$
identifies a transition region that increases with $D$, in a manner
consistent with the prediction of
Eq. (\ref{ginzburg}). By taking the  value of $C u_{t}^{-1}
\approx 0.219$, Eq. (\ref{ginzburg})
gives a size for this transition region
extending up to $q_i \approx
1.25$,$1.27$, and $1.30$, for $D/\Delta x =0.05$, $0.1$, and $0.2$,
respectively, which
is consistent with the trend of the data in Fig.~\ref{master}.
A more detailed comparison of Eq. (\ref{ginzburg})
with our numerical results has not been attempted since Eq. (\ref{ginzburg})
only gives an estimate with undetermined coefficients.

In contrast with the
prediction of linear theory (Eq. (\ref{qM})), the
simulation points for low enough $q_i$ lie on
straight lines, so that the exponent $1/2$ in Eq. (\ref{qM})
is changed by fluctuations to a value close to $1$.
We have defined an effective stability boundary $\tilde{q}_{E}$, a function of
$D$, that separates stable from unstable states within our simulation time
($ t \approx 100$, see Fig. 1).
We define it to be equal to the intersection point between the straight lines
in Fig. 1 and the line $q_{i} = q_{f}$.
This gives $\tilde{q}_{E} = 1.16, 1.14$
and $1.11$ for $D/ \Delta x = 0.05, 0.1$ and $0.2$ respectively.
In addition, our data
disagree with standard criteria for pattern selection according to which
$q_{f} = q_{m}$ or $q_{f} = q_{min}$, when $q_{i}$ lies in the transition
region.

Lastly we describe our numerical results for the temporal evolution of the
dominant periodicity of the configuration $q(t)$ after it becomes linearly
unstable. The function $q(t)$ is defined as $q(t) = \pi n(t)$, where $n(t)$
is the number of nodes of $\psi (x,t)$ per unit length, averaged over
independent runs.
The decay is found to take place in a characteristic time $\tau$
with an apparent divergence as
$q_{i}$ approaches $\tilde{q}_{E}$. A similar behavior has been observed in
simpler zero dimensional stochastic
models used to describe the region that
separates a metastable from an unstable state \cite{re:co89}. There, it is
possible to characterize the temporal evolution in terms of scaling
relations. With this motivation in mind, we have found that our data is
well described by a scaling relation of the form,
\begin{equation}
\label{scaling}
q(t) - \tilde{q}_{E} =
(q_I - \tilde q_E) f \left[ t (q_{i} - \tilde{q}_{E})^z \right],
\end{equation}
as shown in Fig. 2 for $D / \Delta x = 0.1$.
The best scaling is found for $z = 1.7$.
Scaling is also observed for other values of $D$, with $z$ in the range
1.65 - 1.75.

Detailed comparison of our results with experiments would certainly require at
least the
consideration of a two-dimensional sSH equation, but we expect that
the general ideas exposed here, namely the existence of a transition region
separating
stable and unstable states, and the identification of an observable stability
limit shifted with respect to $q_{E}$ and related to an
increase of the characteristic relaxation times, are of wider applicability.
We note, for example, qualitative similarities with some experimental
findings. Lowe and Gollub \cite{re:lo85} observed the decay of states in the
deterministically stable
range and their evolution towards states of wavenumber intermediate
between $q_m$ and $q_{min}$. In addition, in the experiments by Rasenat {\it et
al.} \cite{rasenat}, the Eckhaus boundary was determined
in a way very similar to ours, and it was noted that, in contrast with
deterministic theory, the experimental results for $q_f-q_i$, and the
characteristic time
scales of the decay of the unstable state as a function of $q_i$
were well fitted by straight lines. Further experimental work,
especially in systems in which the intensity of the thermal fluctuations
is relatively large,
is certainly needed to clarify whether the effects described above can be
explained by our theoretical arguments.

This work has been supported by NATO, within the program \lq\lq Chaos, order
and patterns; aspects
on nonlinearity", project number 890482, by the Supercomputer
Computations Research Institute,
which is partially funded by the U.S. Department of Energy
contract No. DE-FC05-85ER25000, and by the Direcci\'on General de
Investigaci\'on Cient\'\i fica y T\'ecnica, contract number PB 89-0424.
The calculations reported here were
performed on the 64k-node Connection Machine 2 at SCRI.

\newpage

\figure{\label{master}} Final average wavenumber $q_{f}$
as a function of the initial wavenumber
$q_{i}$, for several values of the intensity of the fluctuations
$D / \Delta x$:
$\Diamond$, 0.05; $\Delta$, 0.1; and, $\Box$, 0.2. The straight solid lines
are linear fits to the data. The dashed line is $q_{m}$, the wavenumber
corresponding
to the eigenfunction of fastest growth given by linear theory.
The error
bars in the figure represent the $2 \sigma$ confidence interval for the
sample studied.

\figure{\label{figscaling}} Scaled average periodicity of the configuration
as a function of time after the instability (the unscaled values are shown in
the inset) for $D/\Delta x = 0.1$. The value of $z=1.7$ has been used.
The values of $q_{i}$ shown
are (from top to bottom in the right of the inset, and the corresponding
symbols in the
figure), $q_{i} = 1.119,1.211,1.219,1.230,1.238$ and 1.250.


\begin{references}
\bibitem[18]{re:fo1} For $D > 0$ the asymptotic configurations given by
Eq. (\ref{eq:sh}) are not strictly periodic, but have an almost constant
number of nodes (see Refs. \cite{estella,ours}).
\end{references}
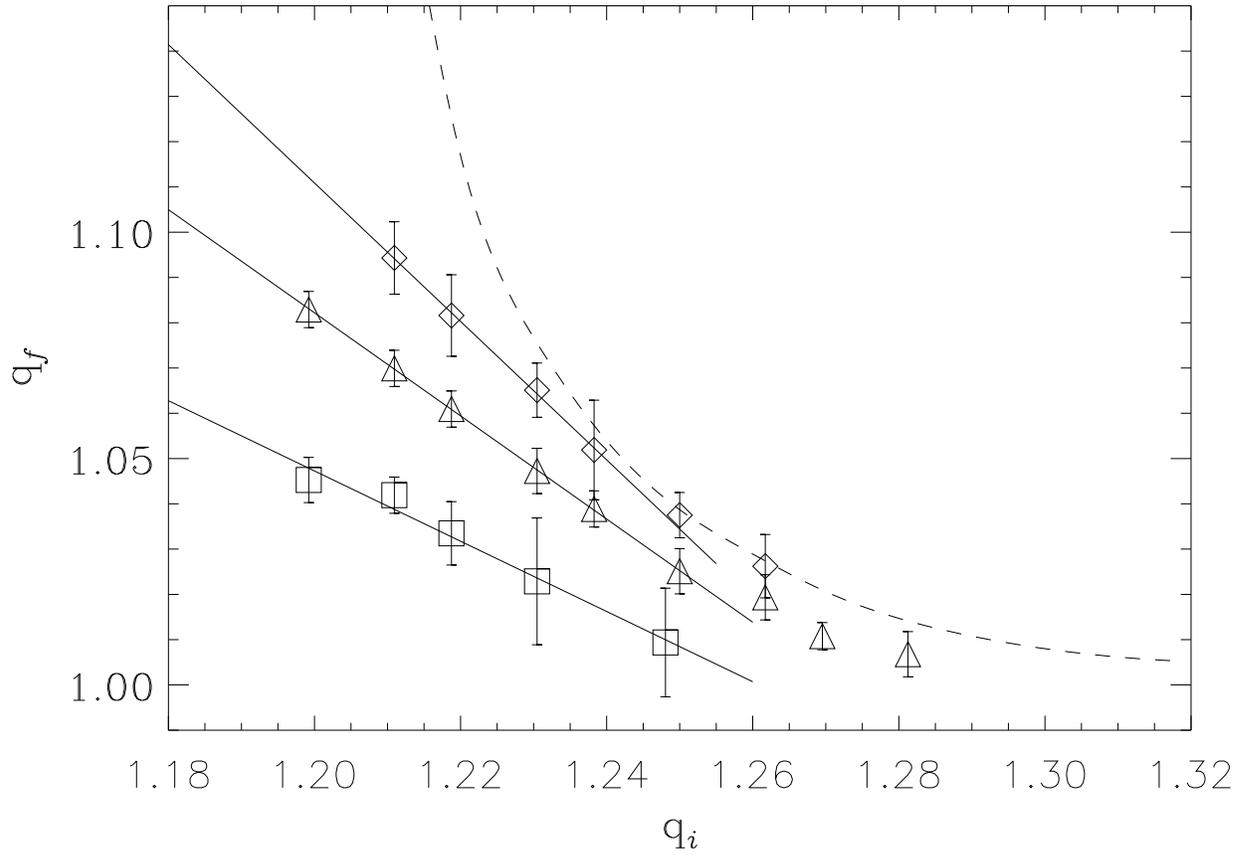
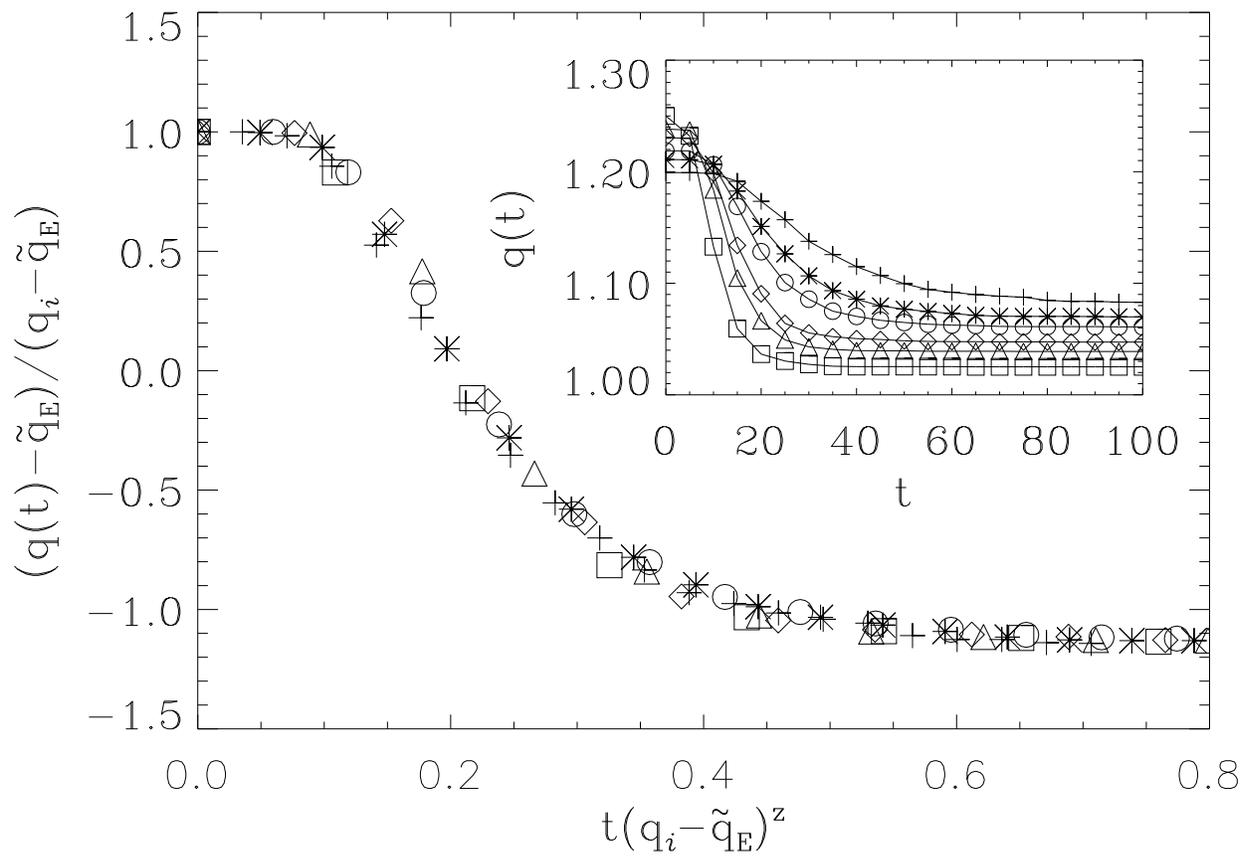
\end{document}